# An Efficient Synchronous Static Memory design for Embedded System

Ravi Khatwal
Research Scholar,
Dept. Of Computer science
Mohan LaL Sukhadia University,
Udaipur, India,

Manoj Kumar Jain
Associate Professor,
Dept. Of Computer science
Mohan LaL Sukhadia University,
Udaipur, India,

## ABSTRACT
Custom memory organization are challenging task in the area of VLSI design. This study aims to design high speed and low power consumption memory for embedded system. Synchronous SRAM has been proposed and analyzed using various simulators. Xilinx simulator simulates the Synchronous SRAM memories which can perform efficient read/write capability for embedded systems. Xinix tool also provide the access time that required selecting a word and reading it. Synchronous Static RAM which has easily read /writes capability and performs scheduled read /writes operation in efficient manner.

## 1. INTRODUCTION
Today memory system is the key component for high performance embedded system. The total memory capacity of a computer system can be visualized as being as hierarchy of various components. The memory hierarchy system consists of all storage devices employed in a computer system from the slow but high capacity auxiliary Memory relative faster than main memory to an even smaller and faster cache require accessible to the high speed processing logic. A special very high speed memory called cache is sometimes used to increase the speed of processing by making current program and data available to the CPU logic is usually faster than main memory access time, with the result that are processing speed is limited primarily by speed of main memory.

Custom memory organization are challenging task in the area of embedded system. Efficient refer to as achieving the maximum productivity with minimum wasted effort or expense and provide the best performance with the least waste of time and effort. A small cache between the CPU and main memory whose access time is close to processor logic clock cycle times. The cache is used for storing a segments of programs currently being executed in the CPU and temporary data frequently needed in the present calculation by making program and data available at a rapid rate , it's possible to increase the performance rate of computer system. Synchronous Static RAMs which have efficient read /write capability and provide efficient process of information.

Synchronous SRAM perform schedule read /write operation in accurate manner.

## 2. RELATED WORK
Jain, M. K., Balakrishnan, M. and Kumar, A. [1] used the scheduler based technique for exploring register file size, number of register window and cache configuration in integrated manner. P. R. Panda, N.D. Dutt and A. Nicoulau [2] designed the scratch-pad memory architecture exploration for application specific designs processors and optimization technique for customize embedded system. Custom memory organization can potentially and significant reduce the system cost and yield performance's.Corre, E.,senn, N. Iulin, .Martin, E. [3] introduce memory synthesis in the standard HLS design flow and proposed new approach to take into account the memory architecture and the mapping in the high level synthesis for real time embedded systems.

Hiser, D., Davidson, J. W. and Whalley, D. B. [4] Analysis a fast, accurate technique to estimate application average memory latency a set of memory hierarchy. In their design meet performance and cost constraints, system designer need fat effective methods to determine the cost and relative performance of many different possible memory configuration. Lio, Y. L., Chen, L.C., Li, M.L and Tsay,, R. S. [5] proposed automatic generation approach for a cycle count accurate memory model from the clocked finite state machine of the cycle accurate memory model. Efficient memory timing model is essential to system level simulation. Yang, S., Verbauwhede, I.M. [6] describe an efficient methology for memory analysis and optimization of embedded system design with the goal of reducing memory usage and basic idea of memory usage for any program.

## 3. SYNCHRONOUS STATIC SRAM
Synchronous static SRAM have high synchronous write capability. When WE is high data loaded into the data input into the word selected by the write address for predictable performance, write address and data input must be stable before a low to high transition. Synchronous static SRAM contain efficient write and read capability. When read and write operation is proper schedule then it can reduce the accessing time and increasing the system performance.

The Xilinx tools [7] provide the efficient access time of memory and provide the accurate processing of information.

In simulation process the CPU must provide memory control signals in such a way so as to synchronize its internal clocked operations with the read and write operation of memory show in figure [1].

## 3.1  16-word by 2-bit Static RAM

The design element is a 16-word by 2-bit static random access memory with synchronous write capability show in figure [2]. When the write enable (WE) is Low, transitions on the write clock (WCLK) are ignored and data stored in the RAM is not affected. When WE is High, any positive transition on WCLK loads the data on the data input (D1:D0) into the word selected by the 4-bit address (A3:A0) show in logic table [1]. For predictable performance, address and data inputs must be stable before a Low-to-High WCLK transition. This RAM block assumes an active-High WCLK.

However, WCLK can be active-High or active-Low. Any inverter placed on the WCLK input net is absorbed into the block. The signal output on the data output pins (O1:O0) is the data that is stored in the RAM at the location defined by the values on the address pins. The INIT_xx properties to specify the initial contents of a wide RAM. INIT_00 initializes the RAM cells corresponding to the O0 output; INIT_01 initializes the cells corresponding to the O1 output, and INPUT BUFFER, OUT BUFFER analysis show in the figure [3, 4].

**Table 1. 16-word by 2-bit static random access memory**

| Inputs | | | Outputs |
|---|---|---|---|
| WE (mode) | WCLK | D1:D0 | O1:O0 |
| 0 (read) | X | X | Data |
| 1(read) | 0 | X | Data |
| 1(read) | 1 | X | Data |
| 1(write) | ↑ | D1:D0 | D1:D0 |
| 1(read) | ↓ | X | Data |
| Data = word addressed by bits A3:A0 | | | |

## 4. SYNTHESIZE DESIGN OF SYNCHRONOUS MEMORY

Static SRAM contain two cross-coupled inverter and varies various transistors. The access transistors are connected to the word line at their respective gate terminals, and the bit line at their source/drain terminals. The word line is used to select the cell while the bit lines are used to perform read/ write operation on the cell. Internally cell holds the stored value on one side and its complements on the other side. A different cell design that eliminates the above limitations is the use of CMOS flip-flop show in figure [5]. The load is replaced by a PMOS transistor. Synchronous CMOS simulation analysis design show in figure [6] and pin configuration show in figure [7]. When memory is read or written, this is called a memory access. A specific procedure is used to control each access to memory, which consists of having the memory controller generate the correct signals to specify which memory location needs to be accessed, and then having the data show up on the data bus to be read by the processor.

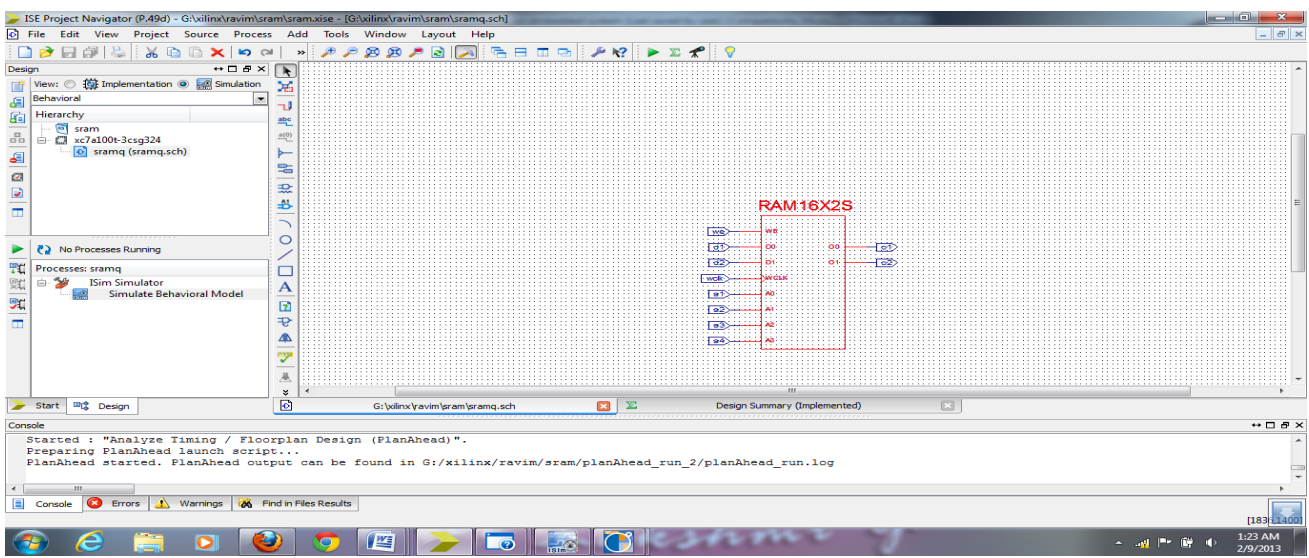

**Fig 1: 16-word by 2-bit static random access memory**

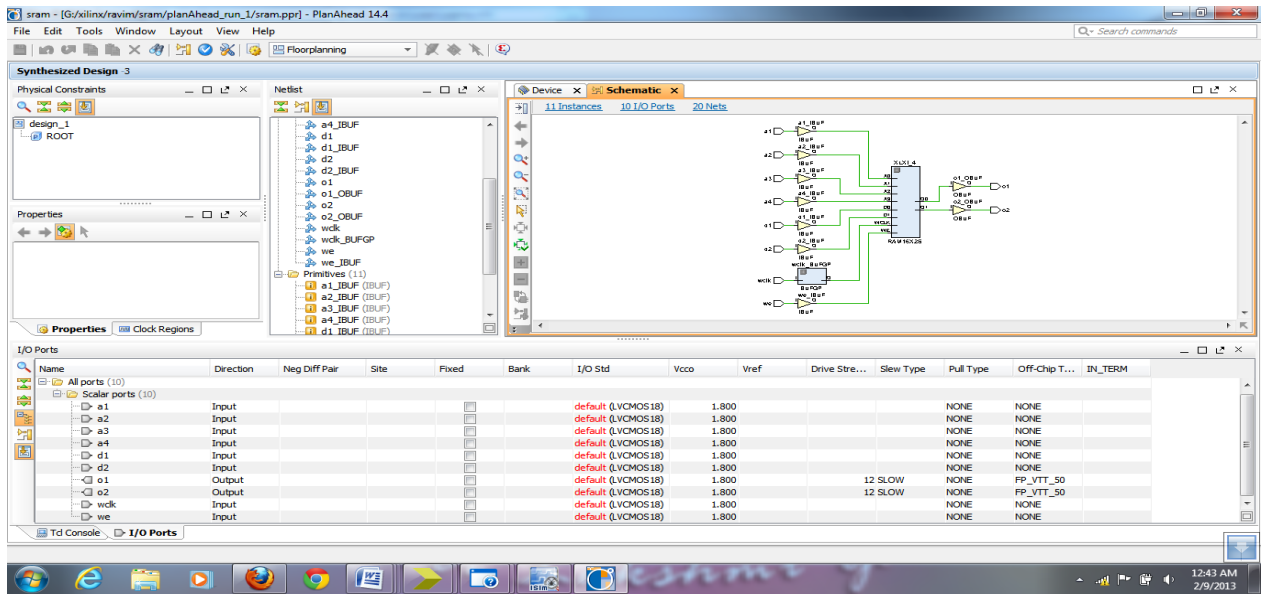

**Fig 2. Architecture design of 16-word by 2-bit static random access memory**

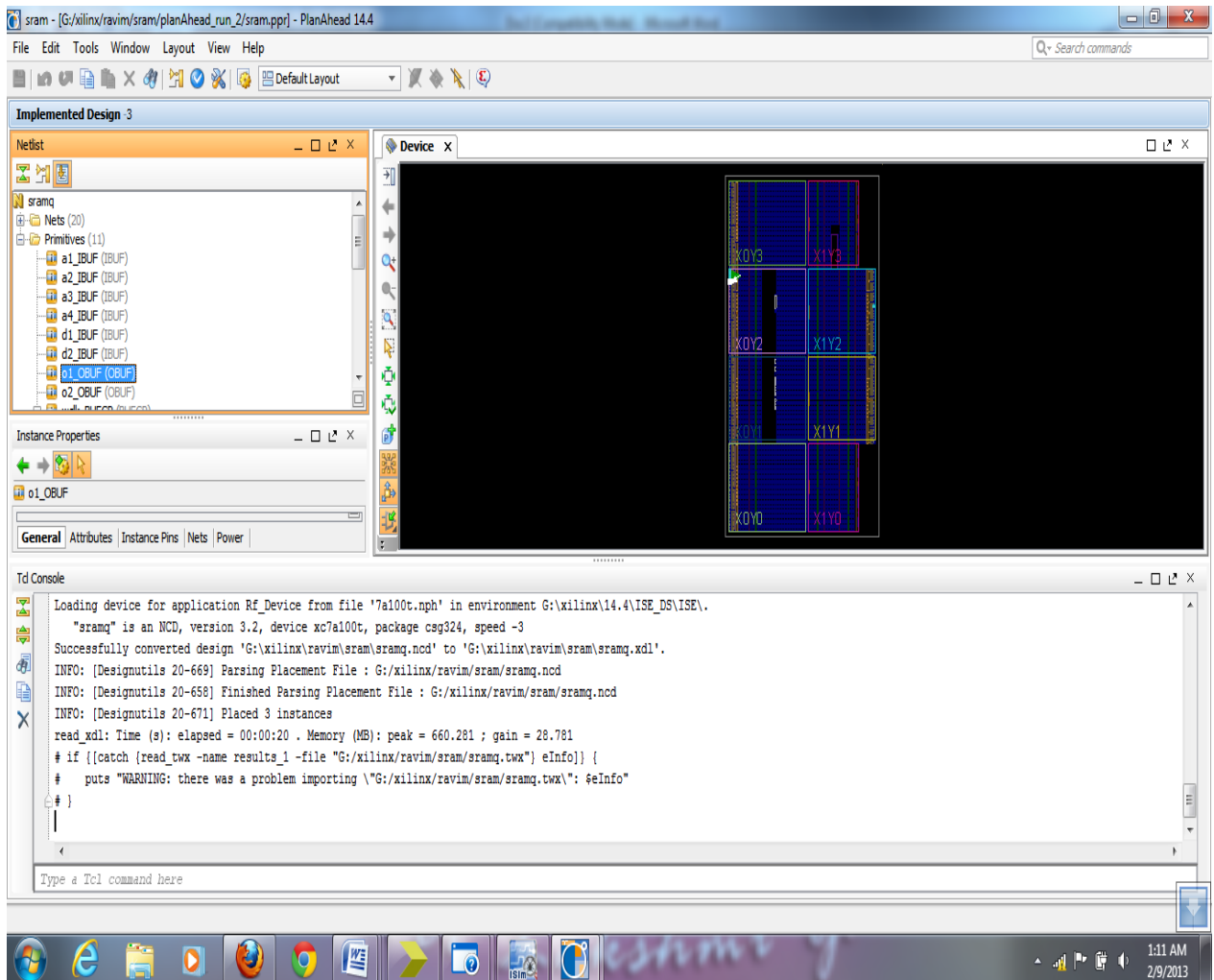

**Fig 3. Input buffer design**

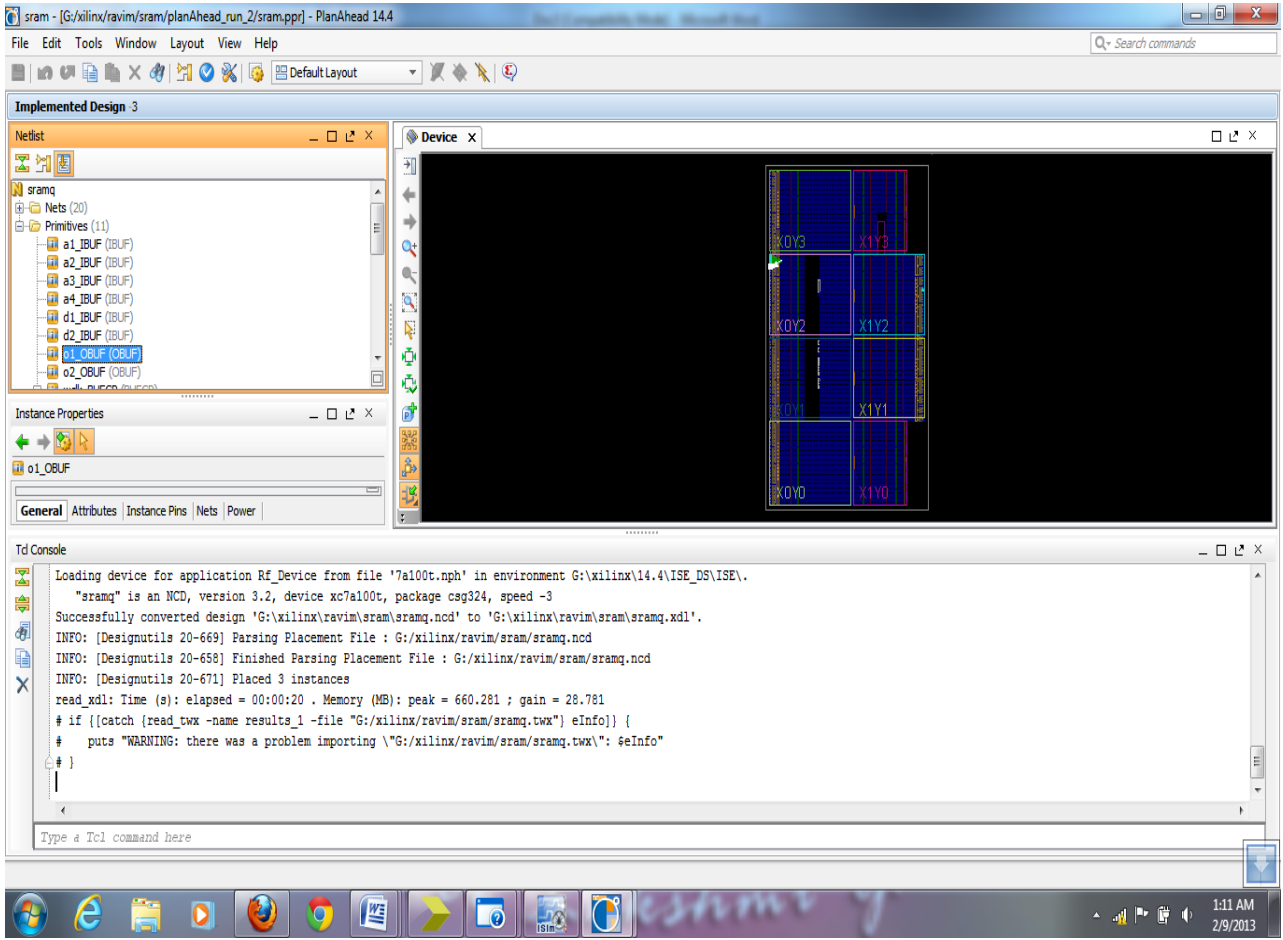

**Fig 4. Output buffer design**

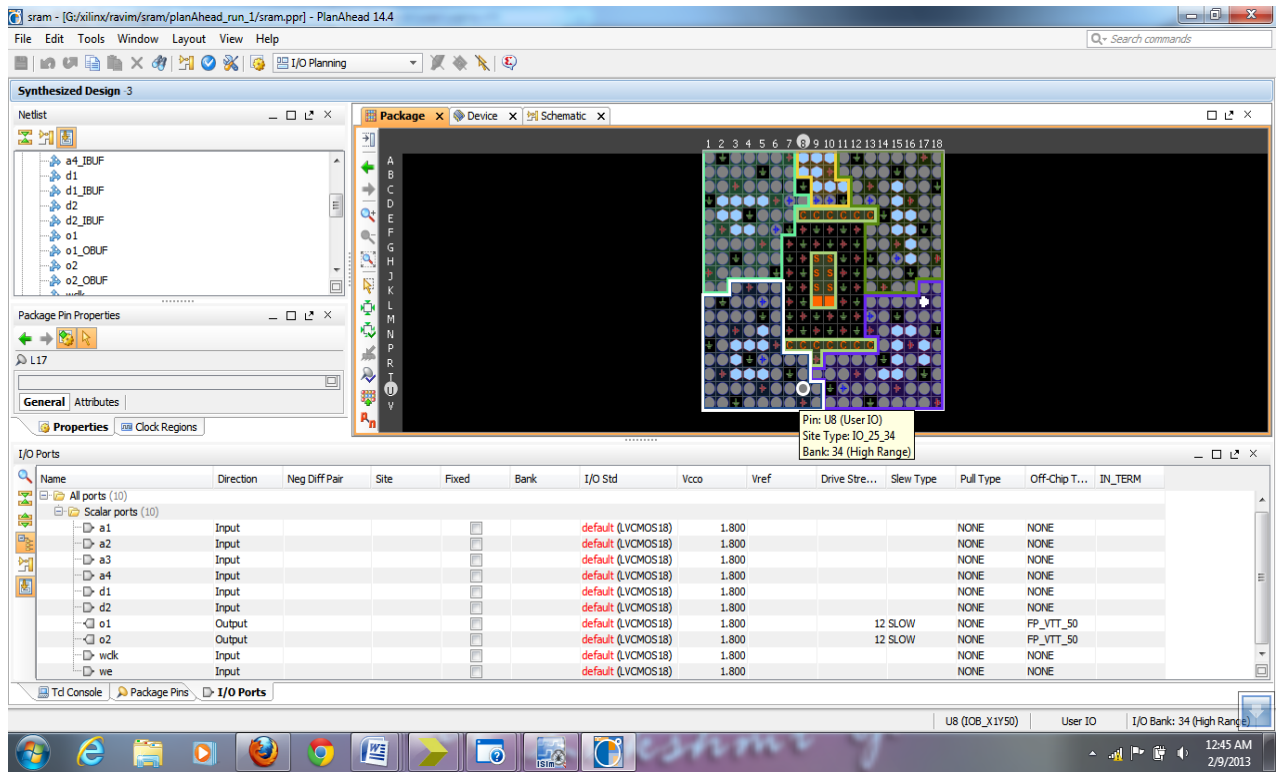

**Fig 5. Synthesize design of synchronous SRAM cell**

**Fig 6. CMOS simulation design of synchronous Static SRAM**

**Fig 7. Pin configuration**

## 5. CLOCK SIGNAL ANALYSIS

The amount of time that it takes for the memory to produce the data required, from the start of the access until when the valid data is available for use, is called the memory's access time, sometimes abbreviated $t_{AC}$. Today's memory normally has access time ranging from 2 to 4 nanoseconds (ns).

Setup and hold times also apply to memory device is loaded into CPU register. The accessing time analysis by the help of 16 x 2s Synchronous static RAM and it's highly efficient because it can reduce the memory access time in efficient manner and Synthesize clock analysis design show in figure [8].

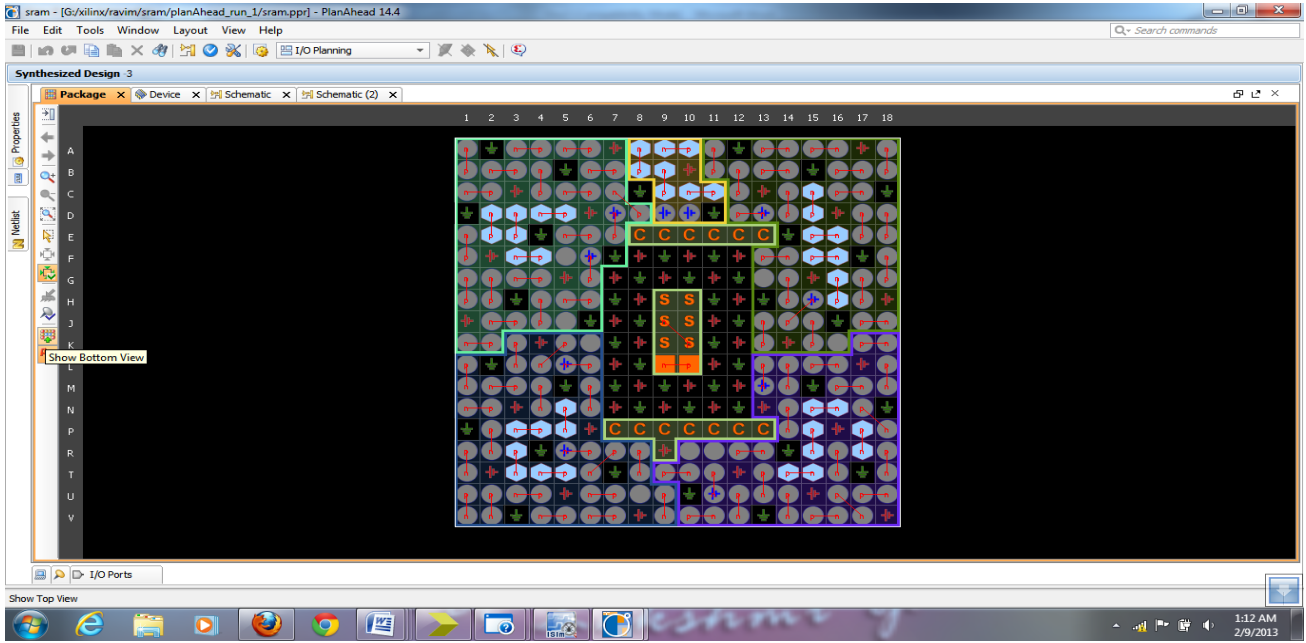

**Fig 8. Clock signal analysis**

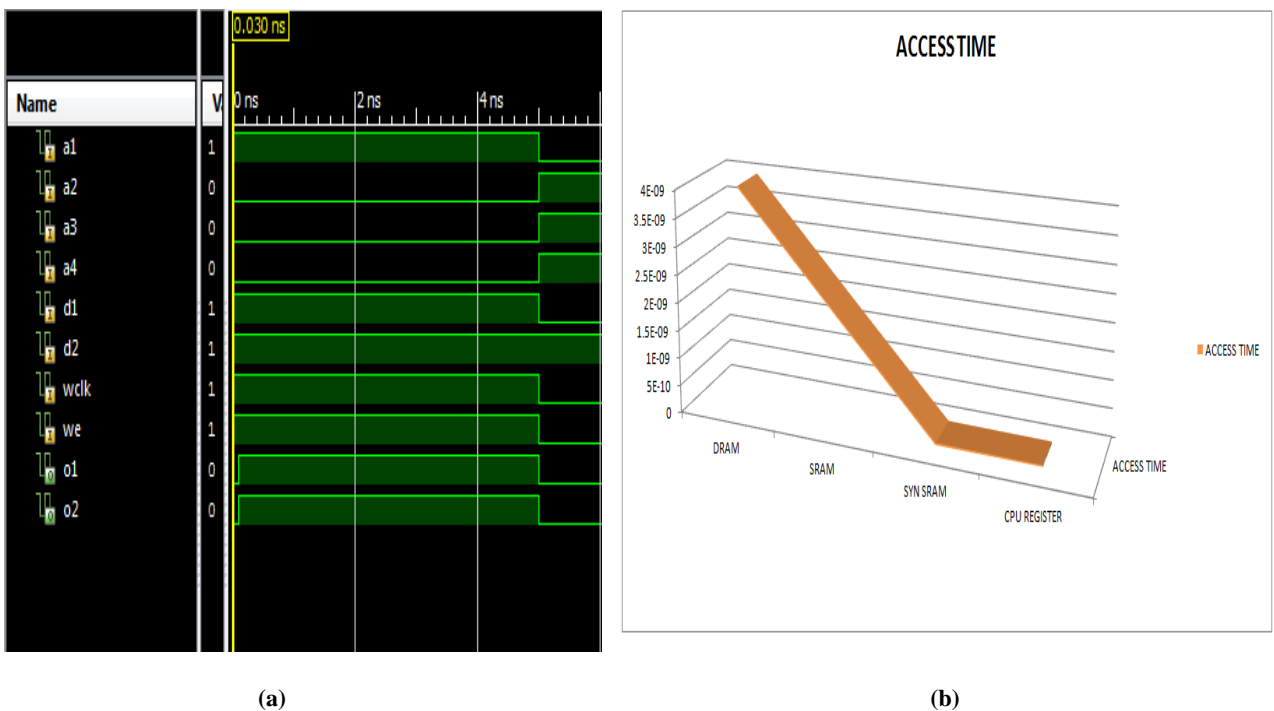

(a)                  (b)

**Fig 9. (a) Synchronous SRAM Simulation result, (b) Analysis of Memory access time**

## 6. EXPERIMENTAL RESULT ANALYSIS

The most common guaranteed response is the propagation delay which is the maximum delay between a change in the input and the correct value appearing in the output. Guaranteed response during a memory device read cycle is the access time. The most common timing requirement are the setup and hold time which are the minimize duration that the data input to a flip-flops has to be at the desired value before and after the relevant clock edge. Synchronous SRAM have their read or write cycles synchronized with the microprocessor clock and therefore can be used in very high speed applications. It can reduce the memory access time and result show in figure [9(a) and 9(b)].

## 7. CONCLUSION AND FUTURE WORK

Synchronous Static random access memory performs better as per simulation analysis and performs efficient write capability for embedded system. The study also revels the fact that data can be easily and quickly accessible in synchronous random access memory. The key idea is to reduce the memory access time and perform the processing of information in efficient manner.

## AUTHOR'S PROFILE

Ravi Khatwal: He is a research scholar in Department of Computer Science, MLSU, Udaipur, and Rajasthan. His research area is VLSI design.

Dr. M.K.Jain: He is Associate Professor in Computer Science at M.L. Sukhadia University Udaipur. His current research interests include Application specific instruction set processor design, wireless sensor networks and embedded systems.